\documentclass[prl,twocolumn,showpacs,preprintnumbers,amsmath,amssymb]{revtex4-2}
\usepackage{amsmath,amssymb,amsthm,dsfont,hyperref,ctable,url,commath}
\usepackage{graphicx}
\usepackage{xcolor}
\newcommand{\ket}[1]{\left | #1 \right\rangle}
\newcommand{\bra}[1]{\left \langle #1 \right |}

\begin{document}

\title{Temporal teleportation with pseudo-density operators:\\ how dynamics emerges from temporal entanglement. }

\author{Chiara Marletto}
\email{chiara.marletto@gmail.com}
\affiliation{Clarendon Laboratory, University of Oxford, Parks Road, Oxford OX1 3PU, United Kingdom, \\Centre for Quantum Technologies, National University of Singapore, 3 Science Drive 2, Singapore 117543, and\\ISI Foundation, Via Chisola, 5, 10126 Torino TO, Italy.}

\author{Vlatko Vedral}
\affiliation{Clarendon Laboratory, University of Oxford, Parks Road, Oxford OX1 3PU, United Kingdom, \\Centre for Quantum Technologies, National University of Singapore, 3 Science Drive 2, Singapore 117543, \\
Department of Physics, National University of Singapore, 2 Science Drive 3, Singapore 117542, and \\ISI Foundation, Via Chisola, 5, 10126 Torino TO, Italy.}

\author{Salvatore Virz\`{i}}
\affiliation{INRIM, strada delle Cacce 91, I-10135 Torino, Italy}

\author{Alessio Avella}

\author{Fabrizio Piacentini}

\author{Marco Gramegna}
\affiliation{INRIM, strada delle Cacce 91, I-10135 Torino, Italy}

\author{Ivo Pietro Degiovanni}

\author{Marco Genovese}
\affiliation{INRIM, strada delle Cacce 91, I-10135 Torino, Italy}
\affiliation{INFN, via P. Giuria 1, I-10125 Torino, Italy}

\date{\today}

\begin{abstract}
We show that, by utilising temporal quantum correlations as expressed by pseudo-density operators (PDOs), it is possible to recover formally the standard quantum dynamical evolution as a sequence of teleportations in time. We demonstrate that any completely positive evolution can be formally reconstructed by teleportation with different temporally correlated states. This provides a different interpretation of maximally correlated PDOs, as resources to induce quantum time-evolution.
Furthermore, we note that the possibility of this protocol stems from the strict formal correspondence between spatial and temporal entanglement in quantum theory. We proceed to demonstrate experimentally this correspondence, by showing a multipartite violation of generalised temporal and spatial Bell inequalities and verifying agreement with theoretical predictions to a high degree of accuracy, in high-quality photon qubits.
\end{abstract}


\maketitle                           %display desi d

Pseudo-density operators (PDOs) were introduced in \cite{Fitzsimons} in order to express quantum spatial and temporal correlations on an equal footing. In usual quantum theory, quantum states, represented as density matrices, are given at a fixed time and then evolved in time through some completely positive (CP) map \cite{zycz}. This is at odds with relativity, where the line of simultaneity is observer-dependent: it therefore represents a problem that hinders quantization of general relativity \cite{lor}. The PDO formulation seeks to rectify this by representing statistics from events with a unique mathematical object, the pseudo-density operator, irrespective of whether the events are space-like, time-like or light-like. Applications of this powerful logic recently led to an experimental simulation to show that the PDO may be a fruitful mode of description even when it comes to esoteric space-times such as the ones that contain open and closed time-like loops \cite{Genovese}. The interested reader is referred to the articles in \cite{Jansen, Robert} for the most up-do-date results on the PDO formalism (see also \cite{Leggett, Brukner, Horsman} for different approaches to temporal quantum correlations).

Given that PDOs encode both spatial and temporal correlations, a natural question arises as to how quantum dynamics can be phrased within such a formulation. Similarly to the relativistic block universe picture, where all the events are laid out in space-time, there is little place for dynamics here: all that matters are space-time relationships between events, which are all encoded in the pseudo-density operator \cite{Robb}.

Here we introduce a formal procedure to obtain quantum dynamics from the PDO description, by generalising the procedure of quantum teleportation to the time domain. In our approach, the temporal correlations of PDOs can be used as a resource to map any state of a qubit to any other, effectively ``teleporting'' it from one time instant to the next, just like spatial entanglement can be used in entanglement-based quantum computing to teleport any quantum state from any spatial location to any another. Quantum dynamics is formally recovered as a sequence of teleportations in time, given a particular PDO used as a resource. Note that the teleportation in time cannot be physically realised as we imagine it with PDOs, because it would require us to have a projective measurement onto states that are not necessarily positive.
However, this formal procedure gives us an alternative way of interpreting what PDOs are: as resources needed to induce dynamics in a static universe. In this paper, we further note that the possibility of this formal analogy between spatial and temporal teleportation is based on the perfect correspondence between spatial and temporal quantum correlations, which is a fundamental principle of quantum theory. We finally experimentally demonstrate this formal correspondence by violating spatial and temporal generalised Clauser-Horne-Shimony-Holt (CHSH) inequalities, showing their agreement with the theoretical predictions to a high degree of accuracy.\\

\textit{Teleportation in time as a formal procedure to recover quantum dynamics.} The logic of deriving dynamics from a given PDO can be illustrated with a simple example, proceeding in perfect analogy with spatial teleportation. To that end, one needs to introduce a set of temporally maximally correlated pseudo-density matrices, in analogy with the Bell basis, as follows.

First, let us summarise the principles of the PDO formalism. Suppose a single qubit, initially in a maximally mixed state, is then measured at two different times (time $a$ and time $b$). Each measurement is performed in all three complementary bases $X, Y,$ and $Z$ (represented by the usual Pauli operators). The evolution is trivial between the two measurements, i.e. the identity operator.
Suppose now that we would like to write the statistics of the measurement outcomes in the form of an operator, generalising the quantum density operator.
Because the state describing these statistics, as we shall see, is hermitian and unit trace, but not positive, we refer to it as a `pseudo-density operator' \cite{Fitzsimons}.

The state can be represented in the following way:
\begin{equation}
\frac{1}{4} \{I + X_aX_b+Y_aY_b+Z_aZ_b\}\; ,
\end{equation}
where $a$ and $b$ are two distinct subsystem, associated each to a 1-qubit Hilbert space, and represent two different times. This operator looks very much like the density operator describing a singlet state of two qubits, however, the correlations all have a positive sign (whereas for the singlet state they are all negative, $\langle X_aX_b\rangle = \langle Y_aY_b\rangle = \langle Z_aZ_b\rangle =-1$). This is a consequence of the fact that it is not a density matrix, because it is not positive (i.e. it has one negative eigenvalue).
We can however trace the label $b$ out and obtain one marginal, i.e. the ``reduced'' state of subsystem $a$. Interestingly, this itself is a valid density matrix (corresponding to the maximally mixed state $I/2$). Likewise for the subsystem $b$.
So, the marginals of this generalised operator are actually both perfectly allowed physical states (just like for a maximally entangled state of two qubits), but the overall state is not (unlike the maximally entangled state of two qubits).

The simple reason why an operator describing temporal correlations cannot always be written as a density matrix is that the outcomes of measurements performed consecutively in the same basis are always perfectly correlated.
That means that we would have the correlation signature of the kind: $\langle X_aX_b\rangle = \langle Y_aY_b\rangle = \langle Z_aZ_b\rangle =1$. However, as we said, there is no allowed density matrix with this signature of correlations: this violates one of the principles of quantum mechanics, because it would require the observables $X_aX_b$, $Y_aY_b$ and $Z_aZ_b$ all to be simultaneously correlated (which is forbidden by complete positivity of the density operator, \cite{Fitzsimons}). Therefore in a pseudo-density operator, although different instances in time can be treated as different subsystems, the price to pay is that the
resulting overall state can have negative eigenvalues (which, therefore, could not be interpreted as probabilities, at least if we think of probabilities either as representing frequencies or degrees of belief. They can also be interpreted as negative probabilities, as already envisaged by Feynman, \cite{FEY}).

There is a set of four maximally correlated PDOs, \cite{Jansen}, which can be considered as a temporal equivalent of the Bell basis:
\begin{eqnarray}
R_{ab}^{(1)} = \frac{1}{4} \{I + X_aX_b+Y_aY_b+Z_aZ_b\}\; \\
R_{ab}^{(2)} = \frac{1}{4} \{I + X_aX_b-Y_aY_b-Z_aZ_b\}\; \\
R_{ab}^{(3)} = \frac{1}{4} \{I - X_aX_b+Y_aY_b-Z_aZ_b\}\; \\
R_{ab}^{(4)} = \frac{1}{4} \{I - X_aX_b-Y_aY_b+Z_aZ_b\}\; .
\end{eqnarray}
These PDOs are ``orthogonal'', in the sense that ${\rm Tr}\{R_{ab}^{(\alpha)}R_{ab}^{(\beta)}\}=\delta_{\alpha\beta}$. We shall now use them to reproduce the teleportation protocol, in the time domain.

We proceed to demonstrate how the general temporal evolution of a qubit from one state (at time $t_a$) to another (at time $t_b$), given by some map $\Phi (\rho)$, can be formally represented as teleportation in time, using (say) a maximally correlated PDO as a resource. We will need three subsystems, labelled as $t_{a}$, $A$ and $t_{b}$; the intermediary subsystem $A$ is an ancilla qubit that, in analogy with spatial teleportation, is formally needed to aid the temporal teleportation from $t_a$ to $t_b$. First, imagine that $\Phi$ is the identity channel and that the initial state of the qubit (to be evolving in time) is $\rho_{t_a} = 1/2 (I_{t_a} +r_x X_{t_a}+r_yY_{t_a} + r_z Z_{t_a})$.
Then we note the following formal identity:
\begin{equation}
{\rm Tr}_{t_aA} \left((R_{t_aA}^{(1)} \otimes I_{t_b}) (\rho_{t_a}\otimes R_{At_b}^{(1)})\right) = \rho_{t_b} = $$$$ = 1/2 (I_{t_b} +r_x X_{t_b}+r_y Y_{t_b} + r_z Z_{t_b})\;.
\end{equation}
This identity is formally equivalent to that underlying standard teleportation; in this case, it is to be interpreted as teleportation in time from instant $t_a$ to instant $t_b$, which describes the evolution of a qubit from a state $\rho_{t_a}$ (at time $t_a$) to the state $\rho_{t_b}=\Phi(\rho_{t_a})$ (at time $t_b$). The maximally temporally correlated PDO $R_{At_b}^{(1)}$ is necessary to achieve teleportation, just like a Bell pair is needed in the spatial case. The subsystems $t_a$ and $A$ are now formally ``projected'' onto the temporally maximally correlated PDO $R_{t_aA}^{(1)}$ (the formal temporal equivalent of a Bell measurement). The outcome of this projection is the density matrix $\rho_{t_b}$ at instant $t_b$. Therefore, this procedure recovers formally the dynamical evolution where a qubit has evolved through time from the instant $t_a$ to the instant $t_b$ under the identity.

In analogy with the spatial case, one can also wonder about teleportation deploying projections on one of the other three maximally correlated PDOs.
We recall that, in the standard teleportation protocol, Alice performs a projective measurement on two qubits in the Bell basis.
There are four possible outcomes, each of which requires Bob to perform a different operation on his qubit in order to obtain the original state of Alice's.
The maximally entangled temporal states are not physical states, as they are not positive operators. Therefore, in this context the projection onto one of them must be intended exclusively as a formal procedure that does not have a physical implementation.
For example, if we were to project on $R_{t_{a}A}^{(2)}$, we would obtain:
\begin{equation}
{\rm Tr}_{t_{a}A} (R_{t_{a}A}^{(2)} \otimes I_{t_b})(\rho_{t_a}\otimes R_{At_{b}}) = U_x \rho_{t_b} U_x^{-1} = $$$$ = 1/2 (I_{t_b} +r_x X_{t_b}-r_yY_{t_b} - r_z Z_{t_b})
\end{equation}
where $U_x$ is a rotation about the x axis; and so on. Therefore, one can interpret projections on different PDOs within the basis $R_{ab}^{(\alpha)}$ as corresponding to the same dynamical evolution, up to a local rotation on the original qubit.\\

What about a more general dynamics? Without loss of generality, we can assume that the density matrix has evolved in the way that the Bloch components $(r_x,r_y,r_z)$ have changed into $(\eta_x r_x,\eta_y r_y, \eta_z r_z)$, where due to the restriction of the complete positivity of the evolution we have $|1\pm \eta_z|\geq |\eta_x\pm \eta_y|$.
To achieve this evolution we need a PDO of the form:
\begin{equation}
R_{At_b} = \frac{1}{4} \{I + \eta_x X_AX_{t_b}+\eta_y Y_AY_{t_b}+\eta_z Z_AZ_{t_b}\}\; .
\end{equation}
This is again in direct analogy with spatial teleportation, where, if a non-maximally entangled channel is used, the teleported state would be related to the original one by a CP map reflecting the non-maximality of the channel, \cite{HOR}.
Therefore, any two-time qubit evolution in time can be formally represented by a teleportation through a suitably chosen PDO that involves three subsystems, as explained. This implies that a two-time dynamical evolution for any system of any dimension can be thus reconstructed, as we can always approximate it arbitrarily well with a sufficient number of qubits, by universality, \cite{DEU}. It is straightforward to extend the procedure to $n$ times, having demonstrated it for two times. Given that the teleportation has occurred between time $t_{n-1}$ and $t_{n}$, it is sufficient to run the same protocol once more, with a new resource PDO $R_{At_{n+1}}$.\\

\textit{Correspondence between spatial and temporal quantum correlations.} The possibility of formally representing dynamics as teleportation in time arises from the formal correspondence between temporal and spatial correlations in quantum mechanics. We would like now to express this correspondence formally, by considering the generalised Bell-type inequalities in space and time. One can first generalise Bell-type inequalities to the temporal domain, \cite{Brukner}. In particular, consider the case of multi-parameter CHSH spatial inequalities -- where, in a bipartite system, Alice and Bob can each choose one of $n$ possible measurement settings (i.e., Boolean observables) $A_1, A_3, ..., A_{2n-1}$ and $B_2, B_4, ..., B_{2n}$. Define the spatial correlation function:

\begin{eqnarray}\label{spaceCHSH}
 S^{(\mathcal{S})}(n)&=&  C^{\mathcal{S}}(A_1,B_2)+C^{\mathcal{S}}(B_2,A_3) +... \nonumber \\
 &+&C^{\mathcal{S}}(A_{2n-1},B_{2n})- C^{\mathcal{S}}(B_{2n},A_1)\;\;\;\;\;
\end{eqnarray}

where $C^{\mathcal{S}}(A, B)$ is the spatial correlation function between two measurement settings $A$ and $B$, chosen as described above.

The multi-parameter CHSH inequality can be written as:
\begin{equation}
S^{(\mathcal{S})}(n)\leq(2n-2)\;.
\end{equation}
In quantum theory, the above inequality is violated - in fact, as $n\rightarrow\infty$, for suitably entangled states the above quantity can be made arbitrarily close to $2n$ \cite{Peres,pr}. The violation of this generalised inequality has been recently demonstrated by highly accurate experiments with photons \cite{Qwiat}.

For the temporal case, one can define an analogous temporal correlation function by considering the observables $A_i$ and $B_i$ as describing two sets of $n$ possible settings, one for each of the two measurements executed in sequence on the same qubit:
\begin{eqnarray}\label{timeCHSH}
 S^{(\mathcal{T})}(n)&=& C^{\mathcal{T}}(A_1,B_2)+C^{\mathcal{T}}(B_2,A_3) +... \nonumber \\
 &+&C^{\mathcal{T}}(A_{2n-1},B_{2n})- C^{\mathcal{T}}(B_{2n},A_1)\;\;\;\;\;
\end{eqnarray}
where, this time, $C^{\mathcal{T}}(A,B)$ is the temporal correlation function between outcomes of observables $A$ and $B$ each measured at two times.

The generalised CHSH inequality in time can be written as:
\begin{equation}
S^{(\mathcal{T})}(n)\leq(2n-2)\;,
\end{equation}
\noindent with a perfect formal parallel with the spatial case. One can show that the above inequality is violated in quantum mechanics. This follows from the fact that the two-point temporal correlation function has the same expression as the spatial two-point correlation when computed for maximally entangled states, \cite{Brukner}.  We can express this with the PDO formalism. Define the temporal average of two observables $A$ (measured at time $t_a$) and $B$ (measured at time $t_b$) as: $\langle\langle A, B\rangle\rangle\equiv{\rm Tr}((A\otimes B) R_{ab})$, where $R_{ab}$ is the relevant pseudo-density operator as defined above.  As $n$ increases, $\sum_{i=1}^{n}\langle\langle A_{2i-1},B_{2i}\rangle\rangle-\langle\langle B_{2n},A_{1}\rangle\rangle$ can be made arbitrarily close to $2n$. Note that one could argue that this fact is not so surprising, because it expresses the well-known fact that measurements cause irreducible perturbations on the quantum state; however, what is interesting is that the way the inequalities are violated by quantum theory is the same in space and time.

This striking correspondence between the CHSH violation in space and time is the key to explain why quantum correlations in space and time can be used in order to achieve, respectively, spatial and temporal teleportation. The former is a well-defined physical protocol, the latter is a formal construction that allows one to reinterpret quantum dynamics as emerging from a time-less PDO.\\
As experimental demonstration of this correspondence, we shall now test the temporal and spatial CHSH inequality violations in photonic systems.\\

\begin{figure}[htbp]
\begin{center}
\includegraphics[width=0.49\textwidth]{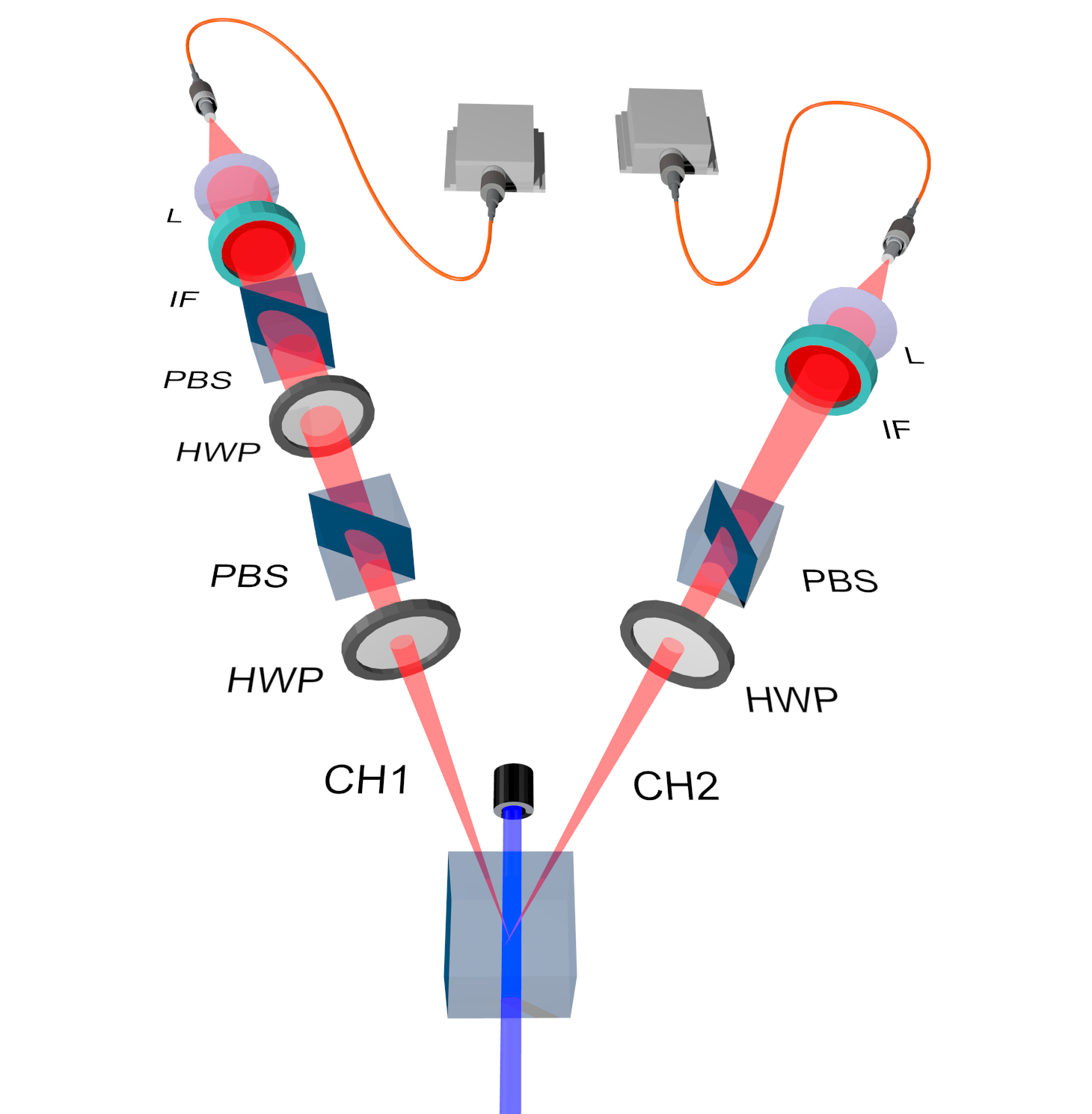}
\caption{Experimental setup. Entangled photons at 808 nm are produced by means of degenerate type-II SPDC, occurring in a BBO crystal pumped by a 76 MHz pulsed laser at 404 nm. Photon on channel 1 (CH1) goes through a double polarization projection stage, each composed of a half-wave plate (HWP) and a polarizing beam splitter (PBS), while photon on channel 2 (CH2) enters a single (identical) stage. After these stages, each photon is addressed to an interference filter (IF) and a lens (L), coupling it into a single-mode fiber feeding a Si-SPAD.
}
\label{setup}
\end{center}
\end{figure}

\begin{figure}[htbp]
\begin{center}
\includegraphics[width=0.49\textwidth]{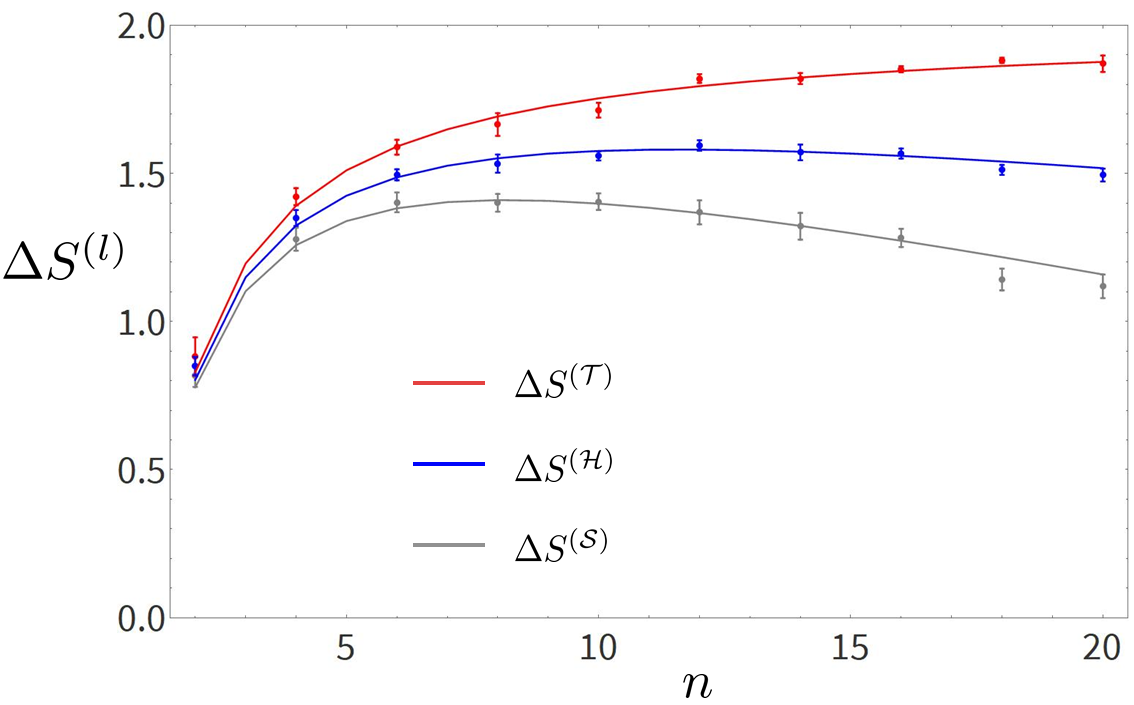}
\caption{Classical bound violation $\Delta S^{(l)}(n)=S^{(l)}(2n)- (2n-2)$ ($l=\mathcal{T},\mathcal{H},\mathcal{S}$) for the multi-parameter CHSH inequalities in the temporal (red), spatial (grey) and ``hybrid'' (blue) domain. The dots represent the experimental results, with the uncertainty bars evaluated as statistical fluctuations among repeated measurement sets, while the solid curves show the theoretically-expected values (for the correlations belonging to the spatial domain, deviations from the ideal case due to the $V_\mathcal{S}=0.982$ estimated visibility of the generated $|\psi_-\rangle$ state were considered).
}
\label{results}
\end{center}
\end{figure}

\textit{The experiment.}
For our experimental demonstration, we exploit the setup shown in Fig. \ref{setup}.
Polarization-entangled photon pairs at 808 nm are produced via degenerate type-II spontaneous parametric down-conversion (SPDC) in a 0.5 mm thick $\beta$-Barium borate (BBO) crystal pumped by a frequency-doubled Ti:Sapphire mode-locked laser (repetition rate: 76 MHz).
The down-converted photons undergo both temporal and phase compensation, and the singlet state $|\psi_-\rangle=\frac{1}{\sqrt2}(|H V\rangle-|V H\rangle)$ is obtained (being $H$ and $V$ the horizontal and vertical polarization components, respectively).
For each entangled pair, the photon on channel 1 (CH1) meets two identical measurement stages in a row, each composed of a half-wave plate (HWP) and a polarizing beam splitter (PBS), while its twin on channel 2 (CH2) undergoes a single polarization measurement, again realized by a HWP followed by a PBS.
After the polarization projections, the photons are spectrally filtered by means of interference filters (IFs, centered onto $\lambda=808$ nm and with a full width at half maximum of 3 nm), coupled to single-mode fibers and addressed to two silicon single-photon avalanche diodes (Si-SPADs), whose output is sent to the coincidence electronics.\\
To evaluate the multi-parameter CHSH inequalities in the spatial domain, for each $n$ value the first measurement stage in CH1 and the one in CH2 realize the set of projections allowing to reach the maximum for $S^{(\mathcal{S})}(n)$, while the second measurement stage on CH1 implements the same projection as the first one, leaving the photon unperturbed.
Concerning the temporal domain, instead, maximal $S^{(\mathcal{T})}(n)$ values are obtained by selecting, for each $n$, the proper projections in the two measurement stages of CH1 (the HWP in the second stage is also responsible for counter-rotating the photon after the first projection). To erase the information on the projection occurred in CH2, we sum the results of two different acquisitions obtained with the CH2 measurement stage realizing orthogonal projections (i.e. $\ket{H}\bra{H}$ and $\ket{V}\bra{V}$).\\
The results of our experiment are reported in Fig. \ref{results}.
%
%For temporal (T, in red) and spatial (S, in grey) domain, the investigated multi-parameter CHSH inequalities have the form of Eq.s (\ref{spaceCHSH}) and (\ref{timeCHSH}), respectively.
%\begin{equation}\label{multiCHSH}
%  S^{(\mathrm{k})}(2n)= \sum_{i=1}^{2n}C^{(\mathrm{k})}(A_{i},B_{i}) + \sum_{i=1}^{2n-1}C^{(\mathrm{k})}(A_{i},B_{i+1})- %  $$$$ + C^{(\mathrm{T})}(A_{2n},B_{1}) \;\;\;\;\; k=\mathrm{T},\mathrm{S}
%\end{equation}
%being $C(\alpha_{i},\beta_{i})$ the correlation coefficient between the two measurement stages, with $\alpha_i$ and $\beta_j$ as respective settings (chosen to grant the maximal violation of the classical bound $S_{C}(N)=N-2$).
In addition to the temporal $(\mathcal{T})$ and spatial $(\mathcal{S})$ multi-parameter CHSH inequalities, we evaluate a third set of inequalities in a sort of ``hybrid'' domain $(\mathcal{H})$, i.e. considering half of the measurements belonging to the temporal domain and half to the spatial one:
\begin{eqnarray}\label{hybridCHSH}
 S^{(\mathcal{H})}(n)&=& C^\mathcal{T}(A_1,B_2)+...+C^\mathcal{T}(B_{n},A_{n+1})+ \nonumber \\
 &+&C^\mathcal{S}(A_{n+1},B_{n+2})+...- C^\mathcal{S}(B_{2n},A_1)\;\;\;\;\;
\end{eqnarray}
%
%\begin{equation}\label{multiCHSHhybrid}
%  S^{(\mathrm{H})}(N)= \sum_{i=1}^{N/2}C^{(\mathrm{S})}(\alpha_{i},\beta_{i}) + $$$$ + \sum_{i=\frac{N}{2}+1}^{N-1}C^{(\mathrm{T})}(\alpha_{i},\beta_{i+1})- C^{(\mathrm{T})}(\alpha_{N},\beta_{1})
%\end{equation}
Figure \ref{results} shows the violations of the classical bound $\Delta S^{(l)}(n)=S^{(l)}(n)-(2n-2)$ obtained for the three cases ($l=\mathcal{S},\mathcal{T},\mathcal{H}$).
For each case, together with the experimental results (dots, with the bars accounting for statistical uncertainties), the expected theoretical behaviour (solid curve) is reported.
While for the temporal domain we consider perfect correlation among measurements, for the spatial one we have to deal with the imperfections of the generated entangled state, inevitably degrading the correlations: for this reason, the theoretical curves were evaluated considering the estimated visibility ($V_{\mathcal{S}}=0.982$) of the realized $|\psi_-\rangle$ state for spatial correlations, and $V_{\mathcal{T}}=1$ for the temporal ones.\\
As evident, the results are in good agreement with the theoretical expectations. The function $S^{(\mathcal{T})}(n)$ keeps growing with $n$, asymptotically reaching the upper limit $2n$, as expected from theory, when $n$ tends to infinity and the angle between measurements settings $A_i$ and $A_{i+1}$ is vanishingly small. In the experiment, from $n=10$ onwards $S^{(\mathcal{S})}(n)$ begins shrinking because of the imperfections of the entanglement produced, becoming more and more relevant as the number of measurements grows.
Obviously, in the hybrid case these imperfections only partially affect the CHSH inequalities, and we obtain a sort of plateau region for $10\leq n\leq16$.\\

\textit{Conclusions.}
We have proposed a scheme to reconstruct quantum dynamics as teleportation in time, using pseudo-density operators. We have also demonstrated experimentally the property that powers this effect, namely the correspondence between spatial and temporal entanglement in quantum theory. We believe that our reformulation of quantum dynamics in terms of PDOs is important from the relativistic perspective too. Quantum field theory has provided a remarkably successful union of quantum physics and special relativity, however, it is fraught with difficulties. Quantum fields are plagued by various divergences but, even more importantly, they are incapable of describing gravity within the same unified framework. One might speculate that this is because space and time are treated as background parameters in quantum field theory, whereas the quantum nature of general relativity might require us to quantise space-time itself (whatever this might mean). It is possible that this would also result in our need to reformulate the core notion of relativity, namely that of causality. Could it be that, at some microscopic scale, the distinction between space-like, time-like and null events evaporates and becomes fuzzy through the application of the quantum superposition principle? PDOs would in that case offer us a way out, since they are a unified way of talking about correlations irrespectively of their origin. The problem of quantising gravity would then become the problem of reconstructing the PDO of the Universe, which would unify not just space and time, but also states and dynamics. Several possible proposals have been put forward to deal with these issues, see e.g. \cite{DH, DH1, DH2, GI}. With our work, we hope to have offered a glimpse of how a possible approach to some of these problems could be, through our theory and experimentation, even though much work clearly needs to be done to complete this vision.\\

\textit{Acknowledgments} \\ VV thanks the Oxford
Martin School, the John Templeton Foundation, the EPSRC
(UK). CM thanks the Templeton World Charity Foundation and the Eutopia Foundation. This research is also supported by the National
Research Foundation, Prime Minister's Office, Singapore, under
its Competitive Research Programme (CRP Award No.
NRF- CRP14-2014-02) and administered by Centre for Quantum
Technologies, National University of Singapore. Furthermore, this research has received funding from PATHOS EU H2020 FET-OPEN grant no. 828946.

%\textit{Author Contribution Statement}\\
%CM and VV proposed the experiment, and are also responsible for the theoretical framework. The experiment was planned by SV, AA, FP, MGr, IPD and MGen (responsible of the lab), and carried on by SV (principal investigator on the experimental side), AA and FP, also responsible for the data elaboration (under the supervision of MGr, IPD and MGen). All authors contributed to the preparation of the manuscript.

\end{document}